\documentclass[pra,aps,amsmath,amssymb,showkeys]{revtex4}

\newcommand{\hh}{{\mathcal{H}}}
\newcommand{\lnp}{{\mathcal{L}}}
\newcommand{\lsa}{{\mathcal{L}}_{s.a.}}
\newcommand{\lsp}{{\mathcal{L}}_{+}}
\newcommand{\bro}{\boldsymbol{\rho}}

\newcommand{\bom}{\boldsymbol{\omega}}

\newcommand{\pen}{\openone}

\newcommand{\am}{{\mathsf{A}}}
\newcommand{\ax}{{\mathsf{X}}}

\newcommand{\az}{{\mathsf{Z}}}

\newcommand{\nm}{{\mathsf{N}}}
\newcommand{\pisf}{{\mathsf{\Lambda}}}
\newcommand{\lasf}{{\mathsf{\Gamma}}}

\newcommand{\km}{{\mathsf{K}}}
\newcommand{\mpi}{{\mathsf{\Pi}}}
\newcommand{\id}{{\mathrm{id}}}

\newcommand{\Tr}{{\mathrm{Tr}}}
\newcommand{\dgt}{{\mathsf{T}}}
\newcommand{\spc}{{\mathrm{spec}}}
\newcommand{\mc}{{\mathcal{M}}}
\newcommand{\nc}{{\mathcal{N}}}

\newcommand{\clb}{{\mathcal{B}}}
\newcommand{\cld}{{\mathcal{S}}}
\newcommand{\rh}{{\mathrm{H}}}
\newcommand{\rrh}{{\mathrm{R}}}
\newcommand{\rfe}{{\mathrm{E}}}
\newcommand{\rff}{{\mathrm{F}}}

\newcommand{\hw}{\widetilde{H}}
\newcommand{\htz}{Z^{\prime}}
\newcommand{\hhx}{\hat{x}}
\newcommand{\hhp}{\hat{p}}

\begin{document}
\clearpage
\preprint{}

\title{R\'{e}nyi and Tsallis formulations of noise-disturbance trade-off relations}

\author{Alexey E. Rastegin}
\affiliation{Department of Theoretical Physics, Irkutsk State University,
Gagarin Bv. 20, Irkutsk 664003, Russia}

\begin{abstract}
We address an information-theoretic approach to noise and
disturbance in quantum measurements. Properties of corresponding
probability distributions are characterized by means of both the
R\'{e}nyi and Tsallis entropies. Related information-theoretic
measures of noise and disturbance are introduced. These
definitions are based on the concept of conditional entropy. To
motivate introduced measures, some important properties of the
conditional R\'{e}nyi and Tsallis entropies are discussed. There
exist several formulations of entropic uncertainty relations for a
pair of observables. Trade-off relations for noise and disturbance
are derived on the base of known formulations of such a kind.
\end{abstract}


\maketitle

\pagenumbering{arabic}
\setcounter{page}{1}

\section{Introduction}\label{sec1}

The Heisenberg uncertainty principle
\cite{heisenberg} is one of the most known restrictions
distinguishing the quantum world from the classical one.
Scientists have made a great effort to understand and extend its
scope and meaning. Basic developments in this direction are
reviewed in \cite{lahti,ww10,brud11}. Various quantitative
measures can be used to describe quantum uncertainties formally
\cite{hall}. In very traditional formulation
\cite{kennard,robert}, we deal with the standard deviations of
corresponding observables. Such an approach was criticized in the
papers \cite{deutsch,maass}, in which entropic formulation has
been developed. The references
\cite{BCCRR10,ccyz12,mdsft13,tomamichel15} considered the entropic
principle in the case of an observer with quantum side
information. An attention is attracted to the entropic formulation
rather due to its connection with some topics of quantum
information theory \cite{ww10,BCCRR10}. On the other hand,
Heisenberg's initial argument is better formulated in terms of
noise and disturbance \cite{ozawa03,ozawa033}. Thus, we cannot
measure precisely an observable without causing a disturbance to
another incompatible observable.

There are more than one approaches to fit a quantitative
formulation of trade-off between noise and disturbance in quantum
measurements. The first universal uncertainty relation of
noise-disturbance type was derived by Ozawa \cite{ozawa033}. Other
formulations have been proposed in
\cite{hall04,werner04,wsu11,blw13,yuoh13,mans14}. The authors of
\cite{kboe14} reported experimental evidences for violation of
so-called Heisenberg's error-disturbance uncertainty relation. For
a discussion of this conclusion, see \cite{blw14} and references
therein. An information-theoretic approach to quantifying noise
and disturbance in quantum measurements has been examined in
\cite{bhow13,cofu13}. Corresponding definitions are based on the
notion of conditional entropy. Formulations of such a kind are
very useful due to several advances. The quantities introduced in
\cite{bhow13} are invariant under relabelling of outcomes. The
possibility of quantum or classical correcting operations is
naturally taken into account. In addition, the
information-theoretic noise can be related to the error
probability of used decision rule.

The present work is devoted to formulating noise-disturbance
relations in terms of generalized entropies. As
information-theoretic measures, entropies of both the R\'{e}nyi
and Tsallis types are used. One of motivations to develop entropic
uncertainty relations is connected with their potential
applications in quantum cryptography \cite{tomamichel11,ngbw12}.
Although R\'{e}nyi's entropies are rather meaningful in studies of
such a kind, the role of Tsallis' ones deserves investigations as
well. Another utility of uncertainty relations with a parametric
dependence was illustrated in \cite{maass}. The presented measures
of noise and disturbance in quantum measurements are defined with
using the conditional R\'{e}nyi and Tsallis entropies. The paper
is organized as follows. Required material is reviewed in Section
\ref{sec2}. First, we discuss quantum measurements and
instruments. Second, basic properties of Tsallis and R\'{e}nyi
entropies are recalled. In particular, we consider relations
between conditional entropies and error probability. Third,
formulations of entropic uncertainty relations for a pair of
observables are discussed. Main results are presented in Section
\ref{sec3}. First, we introduce information-theoretic measures of
noise and disturbance in terms of the conditional Tsallis and
R\'{e}nyi entropies. Reasons for proposed definitions are treated
with the use of essential entropic properties. Using entropic
uncertainty relations, we further derive noise-disturbance
trade-off relations with a parametric dependence. In Section
\ref{sec4}, we conclude the paper with a summary of results.

\section{Preliminaries}\label{sec2}

In this section, preliminary material is reviewed.
First, we recall the formalism of quantum operations, including
quantum measurements and quantum instruments. Second, we write
definitions and some properties of used entropic measures. In
particular, we focus on existing relations between conditional
entropies and error probability. Some formulations of entropic
uncertainty relations are discussed as well.

\subsection{Quantum measurements and instruments}\label{ssc21}

Let $\lnp(\hh)$ be the space of linear operators on
$d$-dimensional Hilbert space $\hh$. By $\lsa(\hh)$ and
$\lsp(\hh)$, we respectively denote the real space of Hermitian
operators on $\hh$ and the set of positive ones. The state of a
quantum system is described by a density matrix $\bro\in\lsp(\hh)$
normalized as $\Tr(\bro)=1$. A common approach to quantum
measurements is based on the notion of positive operator-valued
measures (POVMs). A positive operator-valued measure
$\nc=\{\nm(y)\}$ is a set of elements $\nm(y)\in\lsp(\hh)$
satisfying the completeness relation \cite{peresq}
\begin{equation}
\sum\nolimits_{y}{\nm(y)}=\pen
\ . \label{cmpr}
\end{equation}
Here, the symbol $\pen$ denotes the identity operator on $\hh$. If
the pre-measurement state is described by $\bro$, then the
probability of $y$-th outcome is $\Tr\bigl(\nm(y)\,\bro\bigr)$
\cite{peresq}. The standard measurement of an observable is
described by a projector-valued measure, when POVM elements form
an orthogonal resolution of the identity. As an entropy-based
approach deals with probability distributions, it does not refer
to eigenvalues. Special types of POVM measurements are especially
important. Informationally complete measurements are an
indispensable tool in many questions \cite{prug77,busch91,rbsc04}.
Entropic uncertainty relations for symmetric informationally
complete POVMs are derived in \cite{rastmub}. The informational
power of such preparations and measurements is considered in
\cite{dabo14}.

A unified description of the operation of a laboratory detector is
provided by the concept of quantum instruments \cite{dord13}.
Consider a linear map $\Phi:{\>}\lnp(\hh_{A})\to\lnp(\hh_{B})$.
This map is positive, when $\Phi(\am)\in\lsp(\hh_{B})$ for each
$\am\in\lsp(\hh_{A})$ \cite{nielsen,wilde13}. To describe physical
processes, linear maps must be completely positive
\cite{nielsen,wilde13}. Let $\id_{R}$ be the identity map on
$\lnp(\hh_{R})$, where the space $\hh_{R}$ is assigned to a
reference system. The complete positivity implies that the map
$\Phi\otimes\id_{R}$ with the input space $\hh_{A}\otimes\hh_{R}$
is always positive irrespectively to a dimensionality of
$\hh_{R}$. Any completely positive map can be represented in the
form \cite{nielsen,wilde13}
\begin{equation}
\Phi(\am)=\sum\nolimits_{n}{\km(n){\,}\am{\,}\km(n)^{\dagger}}
\ . \label{osrp}
\end{equation}
Here, the Kraus operators $\km(n)$ map the input space
$\hh_{A}$ to the output space $\hh_{B}$. When physical process is
closed, the corresponding map preserves the trace,
${\Tr}{\bigl(\Phi(\am)\bigr)}=\Tr(\am)$. Trace-preserving
completely positive (TPCP) maps are often called quantum
channels \cite{nielsen,bengtsson}. For a quantum channel, the
Kraus operators satisfy
\begin{equation}
\sum\nolimits_{n}{\km(n)^{\dagger}{\,}\km(n)}=\pen_{A}
\ . \label{clrl}
\end{equation}

Let us consider a collection of completely positive maps
$\mc=\bigl\{\Phi^{(m)}\bigr\}$. The collection $\mc$ is a quantum
instrument, when the maps $\Phi^{(m)}$ are summarized to a
trace-preserving map \cite{bhow13}. For all $\am\in\lnp(\hh_{A})$,
one obeys
\begin{equation}
\sum\nolimits_{m}{\Tr\bigl(\Phi^{(m)}(\am)\bigr)}=\Tr(\am)
\ . \label{clin}
\end{equation}
If the pre-measurement state of an input system is described by
density matrix $\bro$, then the $m$-th outcome occurs with
probability
\begin{equation}
p(m)=\Tr\bigl(\Phi^{(m)}(\bro)\bigr)
\ . \label{prmp}
\end{equation}
In this case, the measuring apparatus will return an output system
in the state described by \cite{bhow13}
\begin{equation}
\bro^{\prime}=p(m)^{-1}\,\Phi^{(m)}(\bro)
\ . \label{prmp1}
\end{equation}
It is convenient to use a trace-preserving completely positive map
defined as
\begin{equation}
\Phi_{\mc}(\bro):=\sum\nolimits_{m}
{\Phi^{(m)}(\bro)\otimes|m\rangle\langle{m}|}
\ . \label{phim}
\end{equation}
The ``flag'' states $|m\rangle$ of an auxiliary system are
orthonormal and, herewith, perfectly distinguishable
\cite{bhow13}. Such states are used for encoding measurements
outcomes.

\subsection{R\'{e}nyi and Tsallis entropies}\label{ssc22}

Together with the Shannon entropy, other entropic
measures are extensively used. Among them, the R\'{e}nyi and
Tsallis entropic functionals are especially important
\cite{bengtsson}. Let discrete random variable $X$ take values on
the finite set $\Omega_{X}$, and let $\{p(x)\}$ be its probability
distribution. For $0<\alpha\neq1$, the R\'{e}nyi entropy is
defined as \cite{renyi61}
\begin{equation}
R_{\alpha}(X):=
\frac{1}{1-\alpha}{\ }{\ln}{\left(\sum_{{\,}x\in\Omega_{X}}{p(x)^{\alpha}}
\right)}
{\,}. \label{renent}
\end{equation}
If the set $\Omega_{X}$ has cardinality $|\Omega_{X}|=d$, then the
maximal value of (\ref{renent}) is equal to $\ln{d}$. It is
reached with the uniform distribution. The entropy (\ref{renent})
is a non-increasing function of order $\alpha$ \cite{renyi61}.
Other properties related to the parametric dependence are
discussed in \cite{zycz}. In the limit $\alpha\to1$, the entropy
(\ref{renent}) gives the Shannon entropy. For $\alpha\in(0,1)$,
the entropy (\ref{renent}) is certainly concave \cite{ja04}.
Convexity properties of $R_{\alpha}(X)$ with orders $\alpha>1$
depend on dimensionality of probabilistic vectors
\cite{bengtsson,ben78}. For instance, for every $\alpha>1$ there
exist an integer $d_{\star}$ such that the entropy (\ref{renent})
is neither convex nor concave for all $d>d_{\star}$ \cite{ben78}.
The two-dimensional case is of special interest. As was explicitly
shown in \cite{ben78}, the binary R\'{e}nyi entropy is concave for
$0<\alpha\leq2$. We also recall that the R\'{e}nyi entropy is
Schur-concave.

Tsallis entropies also form an important family of generalized
entropies. The Tsallis entropy of degree $0<\alpha\neq1$ is
defined as \cite{tsallis}
\begin{equation}
H_{\alpha}(X):=\frac{1}{1-\alpha}{\,}
\left(\sum_{x\in\Omega_{X}}{p(x)^{\alpha}}-1\right)
{\,}. \label{tsent}
\end{equation}
For brevity, we will omit in sums the symbols such as
$\Omega_{X}$. For $0<\alpha\neq1$ and $\xi>0$, we will use the
$\alpha$-logarithm
$\ln_{\alpha}(\xi)=\bigl(\xi^{1-\alpha}-1\bigr)/(1-\alpha)$. One
can rewrite the entropy (\ref{tsent}) as
\begin{equation}
H_{\alpha}(X)=-\sum\nolimits_{x}{p(x)^{\alpha}\,\ln_{\alpha}p(x)}
=\sum\nolimits_{x}{p(x)\,\ln_{\alpha}\!\left(\frac{1}{p(x)}\right)}
{\>}. \label{tsaln}
\end{equation}
When $|\Omega_{X}|=d$, the maximal value of (\ref{tsent}) is equal
to $\ln_{\alpha}(d)$. It is reached with the uniform distribution.
In the limit $\alpha\to1$, we also obtain the
Shannon entropy $H_{1}(X)=-\sum\nolimits_{x}{p(x)\,\ln{p}(x)}$.
Applications of generalized entropies in quantum theory
are reviewed in \cite{bengtsson}. Entropic trade-off
relations for a single quantum channel are discussed in
\cite{rprz12,rast13a}.

In the following, we will also use conditional entropic forms. Let
$Y$ be another random variable. The standard conditional entropy
is defined as \cite{CT91}
\begin{equation}
H_{1}(X|Y):=\sum\nolimits_{y}{p(y)\,H_{1}(X|y)}
=-\sum\nolimits_{x}\sum\nolimits_{y}{p(x,y)\,\ln{p}(x|y)}
\ . \label{cshen}
\end{equation}
Here, we use joint probabilities $p(x,y)$ and the particular
functional
\begin{equation}
H_{1}(X|y)=-\sum\nolimits_{x}{p(x|y)\,\ln{p}(x|y)}
\ , \label{csheny}
\end{equation}
where $p(x|y)=p(x,y)/p(y)$. Similarly to
(\ref{csheny}), we introduce the quantity
\begin{equation}
H_{\alpha}(X|y):=
\frac{1}{1-\alpha}{\,}\left(\sum\nolimits_{x}{p(x|y)^{\alpha}}-1\right)
{\,}. \label{hct00}
\end{equation}

Keeping (\ref{tsaln}) in mind, the two kinds of conditional
Tsallis entropy can be considered \cite{sf06,rastkyb}. These forms
are respectively defined as
\begin{align}
H_{\alpha}(X|Y)&:=\sum\nolimits_{y}{p(y)^{\alpha}\,H_{\alpha}(X|y)}
\ , \label{hct1}\\
\hw_{\alpha}(X|Y)&:=\sum\nolimits_{y}{p(y)\,H_{\alpha}(X|y)}
\ . \label{hct2}
\end{align}
For all $\alpha>0$, the first form (\ref{hct1}) shares the chain
rule \cite{sf06,ZD70}. In this paper, we will rather need another
property. It is natural to demand that conditioning on more may
only reduce the entropy. In effect, the standard conditional
entropy satisfies \cite{CT91}
\begin{equation}
H_{1}(X|Y,Z)\leq{H}_{1}(X|Y)
\ . \label{h12h1}
\end{equation}
For all $\alpha>0$, the second form (\ref{hct2}) of conditional
$\alpha$-entropy obeys \cite{rastit}
\begin{equation}
\hw_{\alpha}(X|Y,Z)\leq\hw_{\alpha}(X|Y)
\ . \label{t12t1}
\end{equation}
The first form (\ref{hct1}) satisfies such a property only for
$\alpha\geq1$ \cite{rastit}. Since the mentioned property is of great
importance in our research, we will use the second form. It should
be noted that the form (\ref{hct2}) does not share the chain rule.
As the first form (\ref{hct1}) of conditional $\alpha$-entropy
obeys the chain rule for all $\alpha>0$ \cite{sf06,ZD70}, it may
be more appropriate in some questions. In the present work,
however, the chain rule is not used.

The R\'{e}nyi case is similar to the Tsallis case in the following
respect. There is no generally accepted approach to the definition
of conditional R\'{e}nyi entropy \cite{tma12}. We will use the
following one. For $0<\alpha\neq1$, the conditional
$\alpha$-entropy is put by \cite{cch97,Kam98,EP04}
\begin{equation}
R_{\alpha}(X|Y):=\sum\nolimits_{y}{p(y)\,R_{\alpha}(X|y)}
\ , \label{rect1}
\end{equation}
where
\begin{equation}
R_{\alpha}(X|y):=
\frac{1}{1-\alpha}{\ }{\ln}{\left(\sum\nolimits_{x}{p(x|y)^{\alpha}}\right)}
{\ }. \label{rect2}
\end{equation}
Like (\ref{renent}), the conditional entropy (\ref{rect1}) is a
non-increasing function of $\alpha$. Another approach for
constructing conditional entropies is connected with the notion of
relative entropy \cite{tomamichel15}. Then conditional entropies
are defined via an optimization problem. The corresponding
formulation of conditional R\'{e}nyi's entropy is considered in
\cite{tomamichel15}, mainly in quantum setting. In the following,
we will use the definition (\ref{rect1}) due to its connection
with error probability.

The limit $\alpha\to\infty$
gives the conditional min-entropy. For the given value $y$, we
define
\begin{equation}
\hhx(y):={\mathrm{Arg}}\max\bigl\{p(x|y):\, x\in\Omega_X\bigr\}
\ . \label{htx12}
\end{equation}
It maximizes $p(x|y)$, i.e.,
$p(x|y)\leq{p}(\hhx|y)$ for all $x\in\Omega_{X}$.
Note that a value (\ref{htx12}) may be not unique. Any of such
values corresponds to the standard decision in the Bayesian
approach \cite{renyi67a}. We then write
\begin{equation}
R_{\infty}(X|y)=-\ln{p}(\hhx|y)
\ . \label{ryin}
\end{equation}
The conditional min-entropy $R_{\infty}(X|Y)$ is defined according
to (\ref{rect1}) and (\ref{ryin}). The following property is
related to conditioning on more. For $0<\alpha\leq1$, the
conditional entropy (\ref{rect1}) satisfies
\begin{equation}
R_{\alpha}(X|Y,Z)
\leq{R}_{\alpha}(X|Y)
\ . \label{rtem2}
\end{equation}
This relation immediately follows from concavity of the entropy
\cite{rastit}. If $|\Omega_{X}|=2$, then the relation
(\ref{rtem2}) is valid for all $\alpha\in(0,2]$. Indeed, the
binary R\'{e}nyi entropy is concave for $0<\alpha\leq2$
\cite{ben78}. Here, the proof holds irrespectively to
dimensionality of any of $Y$ and $Z$. The only restriction is that
the variable $X$ is two-dimensional. With arbitrary finite
$|\Omega_{X}|$, we can use (\ref{rtem2}) only for
$\alpha\in(0,1]$.

The conditional entropy (\ref{rect1}) has interesting properties
and applications in some questions \cite{cch97,Kam98,EP04}. However, this
form does not share the chain rule. Conditional R\'{e}nyi's
entropy can be defined in a way connected with the chain rule
\cite{rw05,gpy09}. In our study, we are rather interested in
properties related to conditioning on more.

\subsection{Relations between conditional entropies and error probability}\label{ssc23}

Although entropic functions are basic measures of
uncertainty, the channel coding theorems are usually stated in
terms of the error probability \cite{CT91}. Hence, relations
between entropies and the error probability are of interest.
Fano's inequality provide an upper bound on the conditional
entropy \cite{FM94}. Known lower bounds on the conditional entropy
are expressed in terms of the error of standard decision. Let
variables $X$ and $Y$ respectively correspond to the input and the
output of a communication system. We should decide on the input
symbols when the output symbols are known. In the standard
decision, we decide in favor of value (\ref{htx12}) for all output
values of $Y$. Then the error probability $\hhp_{e}$ and the
probability of successful estimation $\hhp_{s}$ are written as
\begin{equation}
\hhp_{e}=1-\hhp_{s}
\ , \qquad \hhp_{s}=\sum\nolimits_{y}{p(y)\,p(\hhx|y)}
\ . \label{ppes}
\end{equation}
Due to the Bayesian version of the fundamental Neyman--Pearson
lemma \cite{renyi67a}, no decision can have a smaller error
probability than the standard decision. When there exists a
decision rule with zero error probability, we inevitably have
$\hhp_{e}=0$.

As was shown in \cite{renyi67a,AP67}, the standard conditional
entropy (\ref{cshen}) is bounded from below as
\begin{equation}
-\ln(1-\hhp_{e})\leq{H}_{1}(X|Y)
\ . \label{rlb1}
\end{equation}
This result was extended to some forms of generalized entropies
\cite{rastkyb}. For all $\alpha\in(0,2]$, the conditional entropy
(\ref{hct2}) satisfies
\begin{equation}
\ln_{\alpha}\!\left(\frac{1}{1-\hhp_{e}}\right)\leq\hw_{\alpha}(X|Y)
\ . \label{ttlb2}
\end{equation}
As was recently proved in \cite{rastit}, for $\alpha\in(0,2]$ we
also have
\begin{equation}
2\,\ln_{\alpha}(2){\,}\hhp_{e}\leq\hw_{\alpha}(X|Y)
\ . \label{tbnr1}
\end{equation}
For $\alpha>2$, the lower bound on (\ref{hct2}) depends also on
the dimensionality $d=|\Omega_{X}|$. Namely, we have
\begin{equation}
\frac{d\,\ln_{\alpha}(d)}{d-1}{\ }\hhp_{e}\leq\hw_{\alpha}(X|Y)
\ . \label{ttlb21}
\end{equation}
For all $\alpha\in(0,\infty)$, the conditional R\'{e}nyi entropy
(\ref{rect1}) satisfies
\begin{equation}
-\ln(1-\hhp_{e})\leq{R}_{\alpha}(X|Y)
\ . \label{rrlb1}
\end{equation}
In the binary case, some of the above bounds can be improved
\cite{rastkyb}. For $d=2$, the inequality (\ref{tbnr1}) remains
valid for all $\alpha\in(0,\infty)$. For $d=2$ and
$\alpha\in[1,\infty)$, the conditional R\'{e}nyi entropy
(\ref{rect1}) satisfies
\begin{equation}
2\,\ln_{\alpha}(2){\,}\hhp_{e}\leq{R}_{\alpha}(X|Y)
\ . \label{bnr1}
\end{equation}
For $d=2$ and $\alpha\in(0,1]$, we also have
$(2{\,}\ln{2}){\,}\hhp_{e}\leq{R}_{\alpha}(X|Y)$ \cite{rastkyb}.

Thus, we can claim the following property. If any of the entropies
(\ref{hct2}) and (\ref{rect1}) tends to zero, then $\hhp_{e}$
tends to zero as well. That is, vanishing of conditional entropies
implies that there is a decision function with vanishing error.
In general, this claim is restricted to finite dimensions. For
instance, the bound (\ref{ttlb21}) is applicable only when $d$ is
finite. We will now recall upper bounds related to the
finite-dimensional case.

For an arbitrary decision rule $x^{\prime}=g(y)$, the
corresponding error probability $p_{e}$ is defined similarly to
(\ref{ppes}). The well-known Fano inequality states that
\cite{fano61}
\begin{equation}
H_{1}(X|Y)\leq{h}_{1}(p_{e})+p_{e}\ln(d-1)
\ , \label{fano0}
\end{equation}
where $d=|\Omega_{X}|$ and the binary entropy
$h_{1}(q)=-{\,}q\ln{q}-(1-q)\ln(1-q)$ for $q\in[0,1]$. Let us put
the binary Tsallis entropy
\begin{equation}
h_{\alpha}(q):=-{\,}q^{\alpha}\ln_{\alpha}(q)-(1-q)^{\alpha}\ln_{\alpha}(1-q)
\ . \label{binen}
\end{equation}
As was proved in \cite{rastit}, the conditional entropy
(\ref{hct2}) satisfies
\begin{align}
\hw_{\alpha}(X|Y)&\leq{h}_{\alpha}(p_{e})+p_{e}^{\alpha}\ln_{\alpha}(d-1)
&(0<\alpha<1)
\ . \label{fn01}\\
\hw_{\alpha}(X|Y)&\leq{h}_{\alpha}(p_{e})+p_{e}\ln_{\alpha}(d-1)
&(1<\alpha<\infty)
\ . \label{fn1in}
\end{align}
When $\alpha\to1$, both the formulas (\ref{fn01}) and
(\ref{fn1in}) give the standard Fano inequality (\ref{fano0}).

The authors of \cite{EP04} derived several results concerning the
conditional R\'{e}nyi entropy (\ref{rect1}). For $\alpha\geq1$, the
conditional entropy $R_{\alpha}(X|Y)$ is bounded from above by the
right-hand side of (\ref{fano0}). Indeed, the function
(\ref{rect2}) cannot increase with growing $\alpha$. For
$\alpha\in(1,\infty)$, therefore, we have
$R_{\alpha}(X|Y)\leq{H}_{1}(X|Y)$. Combining this with
(\ref{fano0}) immediately gives the claim. The upper bound
(\ref{fano0}) holds for arbitrary decision rule.

Upper bounds on the conditional R\'{e}nyi entropy of order
$\alpha\in(0,1)$ can be written in terms of the error probability
$\hhp_{e}$ of the standard decision \cite{rastit}. They are based
on one of the results of \cite{ben78}. The conditional R\'{e}nyi
entropy of order $\alpha\in(0,1)$ obeys \cite{rastit}
\begin{equation}
R_{\alpha}(X|Y)\leq\frac{1}{1-\alpha}{\ }
{\ln}{\Bigl((1-\hhp_{e})^{\alpha}+(d-1)^{1-\alpha}\hhp_{e}^{{\,}\alpha}\Bigr)}
\, . \label{rexb1}
\end{equation}
Recall that vanishing of conditional entropies implies that there
is a decision function with zero error probability. On the other
hand, the above bounds of Fano's type imply that conditional
entropies should vanish for $\hhp_{e}\to0$. These results are
essential in motivating measures of information-theoretic noise.
Note that bounds of the Fano type involve dimensionality $d$. We
refrain from discussing relations between conditional entropies
and error probability in the countably-infinite case (see
\cite{hv10} and references therein).

\subsection{General entropic uncertainty relations for finite-level systems}\label{ssc24}

Formulating noise-disturbance relations, we will use
uncertainty relations derived in \cite{zozor13,zozor1311}. For any
$\am\in\lnp(\hh)$, we define $|\am|\in\lsp(\hh)$ to be the
positive square root of $\am^{\dagger}\am$. The singular values
$\sigma_{j}(\am)$ are then introduced as eigenvalues of $|\am|$
\cite{watrous1}. In terms of the singular values, one defines the
Schatten norms widely used in quantum information theory
\cite{watrous1}. We will further use the spectral norm
$\|\am\|_{\infty}=\max\bigl\{\sigma_{j}(\am):\>1\leq{j}\leq{d}\bigr\}$.

Let us consider $d$-dimensional observables $\ax,\az\in\lsa(\hh)$
with the spectral decompositions
\begin{align}
\ax&=\sum_{x\in\spc(\ax)}{x\,\pisf(x)}
\ , \label{xspd}\\
\az&=\sum_{z\in\spc(\az)}{z\,\lasf(z)}
\ . \label{zspd}
\end{align}
Here, the sets $\{\pisf(x)\}$ and $\{\lasf(z)\}$ are corresponding
orthogonal resolutions of the identity. For non-degenerate
observables, we have $\pisf(x)=|{x}\rangle\langle{x}|$ and
$\lasf(z)=|{z}\rangle\langle{z}|$. In this case, the well-known
Maassen--Uffink uncertainty relation \cite{maass} is expressed in
terms of the quantity $c:=\max\bigl|\langle{x}|{z}\rangle\bigr|$.
Inspired by the results of \cite{BCCRR10}, formulations in terms
of quantum conditional entropies were studied. Such uncertainty
relations follow from a few simple properties \cite{ccyz12}
including monotonicity of relative entropies under the action of
TPCP maps. For a wide range of parameter values, this important
fact has been proved for the so-called ``sandwiched'' R\'{e}nyi
entropy. This collection of new relative entropies of R\'{e}nyi's
type was introduced and motivated in \cite{mdsft13}. An
application of such entropies to studying noise-disturbance
trade-off relations may be a theme of separate investigation.

When the range of summation is clear from the context, we will
omit symbols like $\spc(\ax)$ and $\spc(\az)$. The authors of
\cite{zozor1311} have addressed a problem of finding $c$-optimal
bounds on the sum of corresponding entropies. As a measure of
uncertainty in quantum measurements, one uses generalized
entropies of the $(h,\phi)$-form examined in the papers
\cite{sal1,sal2}. We will consider a particular case of this
entropic family. Namely, for any $\alpha>0$ we define
\begin{equation}
E_{\alpha}^{f}(X):=\frac{1}{1-\alpha}{\ }f\!\left(\sum\nolimits_{x}{p(x)^{\alpha}}\right)
\, . \label{faen}
\end{equation}
Here, the function $\xi\mapsto{f}(\xi)$ should be  continuous and
strictly increasing with $f(1)=0$. This choice obeys the
conditions required in \cite{zozor1311} and is completely
sufficient for our purposes. Indeed, the R\'{e}nyi entropy
(\ref{renent}) and the Tsallis entropy (\ref{tsent}) are
respectively obtained from (\ref{faen}) with particular choices
\begin{equation}
f^{(R)}(\xi):=\ln\xi
\ , \qquad
f^{(T)}(\xi):=\xi-1
\ . \label{rtpc}
\end{equation}
We avoid considering entropies of more general kind, since our
constructions will involve conditional entropies.

Measuring the observable $\ax$ in the pre-measurement state $\bro$,
the outcome $x$ occurs with the probability
$\Tr(\pisf(x)\,\bro)$. Substituting this distribution into
(\ref{faen}), we obtain the quantity
\begin{equation}
\rfe_{\alpha}^{f}(\ax;\bro)=
\frac{1}{1-\alpha}{\ }
f\!\left(\sum\nolimits_{x}{\bigl[\Tr\bigl(\pisf(x)\,\bro\bigr)\bigr]^{\alpha}}\right)
\, . \label{faex}
\end{equation}
This quantity characterizes an amount of uncertainty in performed
quantum measurement. In the case of POVM $\nc=\{\nm(y)\}$, the
entropy $\rfe_{\alpha}^{f}(\nc;\bro)$ is given similarly to
(\ref{faex}), but with the probabilities
$\Tr\bigl(\nm(y)\,\bro\bigr)$.

To two observables $\ax,\az\in\lsa(\hh)$, we assign
the characteristic
\begin{equation}
c:={\max}{\Bigl\{
\|\pisf(x)\,\lasf(z)\|_{\infty}
:{\>}
x\in\spc(\ax),{\>}z\in\spc(\az)
\Bigr\}}
{\,}, \label{cxzp}
\end{equation}
and related parameter $\eta:=\arccos{c}$. Concerning (\ref{cxzp}),
the following fact should be noticed. It is easy to see that
$\|\am^{\dagger}\|_{\infty}=\|\am\|_{\infty}$ for any $\am$. Since
both the projectors $\pisf(x)$ and $\lasf(z)$ are Hermitian, we
then get
\begin{equation}
\|\pisf(x)\,\lasf(z)\|_{\infty}=\|\lasf(z)\,\pisf(x)\|_{\infty}
\ . \label{etcs}
\end{equation}
For non-degenerate observables, the characteristic (\ref{cxzp}) is
reduced to the maximal overlap between eigenstates of $\ax$ and
$\az$, i.e., to
$\max\bigl|\langle{x}|{z}\rangle\bigr|$. As follows from the
unitarity, the latter ranges between $d^{-1/2}$ and $1$.
Introducing the parametric sum
\begin{equation}
\cld_{\alpha}(\theta):=
\left\lfloor\frac{1}{\cos^{2}\theta}\right\rfloor
(\cos^{2}\theta)^{\alpha}+
\left(
1-\left\lfloor\frac{1}{\cos^{2}\theta}\right\rfloor
\cos^{2}\theta
\right)^{\!\alpha}
, \label{sfth}
\end{equation}
for all $\alpha,\beta\geq0$ we define the quantity
\begin{equation}
\overline{\clb}_{\alpha,\beta;f}(c):=
\underset{\theta\in[0,\eta]}{\min}
{\left(
\frac{{f}{\bigl(\cld_{\alpha}(\theta)\bigr)}}{1-\alpha}
+\frac{{f}{\bigl(\cld_{\beta}(\eta-\theta)\bigr)}}{1-\beta}
\right)}
{\,}. \label{ovbn}
\end{equation}
For all $\alpha,\beta\geq0$ and two finite-dimensional observables,
the corresponding generalized entropies satisfy the
state-independent lower bound \cite{zozor1311}
\begin{equation}
\rfe_{\alpha}^{f}(\ax;\bro)+\rfe_{\beta}^{f}(\az;\bro)\geq
\overline{\clb}_{\alpha,\beta;f}(c)
\end{equation}
This generalized-entropy uncertainty relation for two
observables has been proved recently in
\cite{zozor1311}. Note that our notation slightly differs from the
notation of \cite{zozor1311} in minor respects. Substituting the
functions (\ref{rtpc}), we obtain the lower bounds for both the
Tsallis and R\'{e}nyi formulations
\begin{align}
\overline{\clb}_{\alpha,\beta}^{{\,}(T)}(c)
&:=\underset{\theta\in[0,\eta]}{\min}
{\left(
\frac{\cld_{\alpha}(\theta)-1}{1-\alpha}
+\frac{\cld_{\beta}(\eta-\theta)-1}{1-\beta}
\right)}
{\,}, \label{ovbt}\\
\overline{\clb}_{\alpha,\beta}^{{\,}(R)}(c)
&:=\underset{\theta\in[0,\eta]}{\min}
{\left(
\frac{\ln\cld_{\alpha}(\theta)}{1-\alpha}
+\frac{\ln\cld_{\beta}(\eta-\theta)}{1-\beta}
\right)}
{\,}, \label{ovbr}
\end{align}
where $\eta=\arccos{c}$. In the next, we will use these bounds in
obtaining both the R\'{e}nyi and Tsallis formulations of
noise-disturbance relations. It should be noted that the authors
of \cite{zozor1311} derived their uncertainty relations also for
the case of two POVMs. However, a treatment becomes much
more complicated. In particular, it depends on the maximal
spectral norm among elements of a single POVM. On the other hand,
the results (\ref{ovbt}) and (\ref{ovbr}) for projective
measurements are sufficient for our aims.

We will also use entropic uncertainty relations of the
Maassen--Uffink type. This approach
was developed in deriving uncertainty relations in terms of
R\'{e}nyi \cite{birula06} and Tsallis entropies \cite{rast104}.
Using Riesz's theorem leads to a specific condition imposed on
entropic parameters. Developing this approach in some physical
cases of specific interest is considered in
\cite{rast12num,rast12quasi,bosyk13a}. The corresponding Tsallis
entropies satisfy \cite{rast104}
\begin{equation}
\rh_{\alpha}(\ax;\bro)+\rh_{\beta}(\az;\bro)
\geq{\ln_{\mu}}{\bigl(c^{-2}\bigr)}
{\>}, \label{gntl}
\end{equation}
where $1/\alpha+1/\beta=2$ and $\mu=\max\{\alpha,\beta\}$. Under
the same condition on $\alpha$ and $\beta$, the corresponding
R\'{e}nyi entropies satisfy \cite{rast104}
\begin{equation}
\rrh_{\alpha}(\ax;\bro)+\rrh_{\beta}(\az;\bro)
\geq-2\ln{c}
{\>}. \label{gnrl}
\end{equation}
As was motivated in \cite{zozor1311}, the bounds (\ref{ovbt}) and
(\ref{ovbr}) are not always $c$-optimal. In some cases, bounds of
the Maassen--Uffink type are stronger. Thus, we will also derive
noise-disturbance relations with the use of (\ref{gntl}) and
(\ref{gnrl}). The considered bounds are formulated in terms of only
one quantity (\ref{cxzp}). Another approach to obtaining
entropic bounds is dealing with more matrix elements of the form
$\langle{x}|{z}\rangle$. This important topic has been studied in
recent works \cite{prz13,fgg13,rpz14}. Bounds of such a kind are
not used in the following.

\section{Main results}\label{sec3}

In this section, we formulate noise-disturbance
relations with the use of generalized entropies. First, we use the
conditional entropies (\ref{hct2}) and (\ref{rect1}) to quantify
information-theoretic noise and disturbance in quantum
measurements. The introduced measures are a natural extension of
the quantities proposed in \cite{bhow13}. Second, we derive
nontrivial lower bounds on the sum of introduced measures of
information-theoretic noise and disturbance.

\subsection{Information-theoretic noise and disturbance}\label{ssc31}

Let $\ax$ and $\az$ be observables of a studied quantum
system $A$ with $d$-dimensional state space. It is assumed to be
subjected to a measuring apparatus $\mc$. We consider the
following two variants of correlation experiments performed with
$\mc$ \cite{bhow13}. In the first experiment, some source produces
eigenstates of $\ax$ at random. For non-degenerate $\ax$, it
should produce each eigenstate $|x\rangle$ with the probability
$1/d$. According to (\ref{xspd}), the integer
$d_{x}:=\Tr\bigl(\pisf(x)\bigr)$ gives degeneracy of the
eigenvalue $x$. Therefore, it should be taken at random with the
probability $d_{x}/d$ \cite{bhow13}. The corresponding eigenstate
is written as $\pisf(x)/d_{x}$. We feed each of the eigenstates of
$\ax$ into the apparatus $\mc$ and ask for correlations of the
observed outcomes $m$ with the eigenvalues of $\ax$. The first
experiment focuses on the average performance of the apparatus in
discriminating between possible values of $\ax$. Only the
classical outcomes are used for guessing in the first experiment
\cite{bhow13}.

In the second experiment, another source produces eigenstates of
$\az$ at random. Due to (\ref{zspd}), each eigenvalue $z$ is
associated with the density matrix $\lasf(z)/d_{z}$, where
$d_{z}:=\Tr\bigl(\lasf(z)\bigr)$. The corresponding probability is
given by $d_{z}/d$ including $1/d$ for non-degenerate $\az$. The
eigenstates of $\az$ are fed through the apparatus $\mc$. Then the
task is to guess the input eigenvalue $z$. Contrary to the first
test, we allow an arbitrary operation $\Psi$ acting on both the
classical outcome $m$ and the actual quantum output of the
apparatus. This operation is aimed to reverse a disturbance
generated by $\mc$ during the act of measurement. Thus, the notion
of disturbance is related to the irreversible character of quantum
measurements \cite{bhow13}. The disturbance is zero, whenever the
input of the apparatus can be recast perfectly after the
correction stage. A significance of unavoidable disturbance was
emphasized in \cite{bhow13}.

The pre-measurement state $\bro$ will lead to
statistics with probabilitis $p(x)=\Tr\bigl(\pisf(x)\,\bro\bigr)$.
Measuring by some instrument $\mc$ results in outcomes $m$ with
corresponding probabilities (\ref{prmp}). We wish to estimate
quantitatively, whether the apparatus $\mc$ measures $\ax$
accurately. As the actual measurement outcome is kept, we try to
guess which eigenstate has been input. The guessed value
$x^{\prime}$ is represented as a function $g(m)$ of the
measurement outcome. The ``maximum {\it a posteriori} estimator''
always gives $\hhx$ defined similarly to (\ref{htx12}). Of course,
an optimization over guessing functions can be taken into account.

When the pre-measurement state is taken to be completely mixed
state $\bro_{*}=\pen_{A}/d$, we deal with the probability distribution
$p(x)=d_{x}/d$. For non-degenerate observables, the input random
variable $X$ will be uniformly distributed. In effect, there are
no general reasons to prefer one value of $x$ to another. Then
different outcomes $x$ will equally contribute to an
information-theoretic measure of noise. In the case of degeneracy,
equal weights of the outcomes are rescaled appropriately. Due to
Bayes' rule, the joint probability distribution of random
variables is written as
\begin{equation}
p(m,x)=p(x){\,}p(m|x)=\Tr\bigl(\pisf(x)\,\bro_{*}\bigr){\,}p(m|x)
\ . \label{jpmx}
\end{equation}
The conditional probability $p(m|x)$ is obtained by substituting the
density matrix $\pisf(x)/d_{x}$ into the right-hand side of
(\ref{prmp}). The joint distribution (\ref{jpmx}) describes a
common statistics of the input variable $X$ and the output
variable $M$. Hence, we can obtain conditional probabilities
$p(x|m)=p(m,x)/p(m)$. The idea is that a contribution of the given
$m$ into a measure of noise should depend on corresponding
conditional probabilities $p(x|m)$. The following property is
physically natural for each fixed $m_{\star}$. The closer
distribution $p(x|m_{\star})$ to uniform, the larger its
contribution to a measure of noise.

Using generalized conditional entropies, we will develop the ideas
of \cite{bhow13}. For $\alpha\in(0,1]$, we define R\'{e}nyi's
information-theoretic noise of the instrument $\mc$ as
\begin{equation}
N_{\alpha}^{(R)}(\mc,\ax):=R_{\alpha}(X|M)
\ . \label{renos}
\end{equation}
Here, $R_{\alpha}(X|M)$ is the conditional R\'{e}nyi
$\alpha$-entropy calculated from the joint probability
distribution $p(m,x)$. In the case $d=2$, we allow to
use (\ref{renos}) for $\alpha\in(0,2]$. For all $\alpha>0$, we
define Tsallis' information-theoretic noise as
\begin{equation}
N_{\alpha}^{(T)}(\mc,\ax):=\hw_{\alpha}(X|M)
\ . \label{tsanos}
\end{equation}
The quantities (\ref{renos}) and (\ref{tsanos}) are respectively
R\'{e}nyi's and Tsallis' versions of the information-theoretic
measure introduced in \cite{bhow13}. The latter is obtained from
(\ref{renos}) and (\ref{tsanos}) in the case $\alpha=1$. Note that
the definitions (\ref{renos}) and (\ref{tsanos}) do not assume an
optimization over guessing functions. This question is closely
related to the restriction $\alpha\in(0,1]$ used in the R\'{e}nyi
case. Let $M\mapsto{g}(M)$ be a function of random variable $M$.
The standard conditional entropy obeys
\begin{equation}
{H_{1}}{\bigl(X\big|g(M)\bigr)}\geq{H}_{1}(X|M)
{\ .} \label{stgn}
\end{equation}
Like (\ref{h12h1}), the inequality (\ref{stgn}) is connected with
the concavity property. In a similar manner, for all $\alpha>0$
the conditional entropy (\ref{hct2}) satisfies
\begin{equation}
{\hw_{\alpha}}{\bigl(X\big|g(M)\bigr)}\geq\hw_{\alpha}(X|M)
{\ .} \label{tsgn}
\end{equation}
This result can be proved similarly to (\ref{stgn}). The case of
R\'{e}nyi's entropies is more complicated. Together with (\ref{rtem2}),
for $\alpha\in(0,1]$ we can obtain
\begin{equation}
{R_{\alpha}}{\bigl(X\big|g(M)\bigr)}\geq{R}_{\alpha}(X|M)
{\ .} \label{rngn}
\end{equation}
For orders $\alpha>1$, we cannot assume concavity of
the conditional R\'{e}nyi $\alpha$-entropy. As mentioned in
section 2.3 of \cite{bengtsson}, the R\'{e}nyi $\alpha$-entropy
is not concave for $\alpha>\alpha_{\star}>1$, where
$\alpha_{\star}$ depends on dimensionality of probabilistic vectors.
Unfortunately, sufficiently precise lower bounds on $\alpha_{\star}$
are not known. In principle, for $\alpha>\alpha_{\star}$ we could
rewrite (\ref{renos}) with an optimization over guessing
functions. At the same time, the property (\ref{rtem2}) is crucial
in proving information-theoretic relations for noise and
disturbance. Within the R\'{e}nyi formulation, we therefore focus
on the range $\alpha\in(0,1]$ in a finite-dimensional case and on
the range $\alpha\in(0,2]$ in the two-dimensional case. Finally,
we point out a conclusion based on the formulas of
Subsection \ref{ssc23}. Each of the information-theoretic noise
(\ref{renos}) and (\ref{tsanos}) vanishes, if and only if the
minimal error probability tends to zero.

The above scheme seems to be more natural for non-degenerate
observables, when each outcome $x$ is taken with the probability
$1/d$. The non-degenerate case is not very
restrictive. Of course, physical systems often have degenerate
observables. As a rule, the degeneracy is connected with
symmetries of the system. However, real systems are typically
subjected to some amount, even if small, of disorder. Such small
imperfections will inevitably break the degeneracy. In this sense,
the results for non-degenerate observables are sufficiently
general.

The second question concerns an information-theoretic approach to
quantifying the unavoidable disturbance. To do so, we consider the
second observable $\az$. As mentioned above, the main difference
between the first and the second correlation experiments is that,
in the second one, we permit to use both the classical outcome and
the output quantum system. To fit the unavoidable disturbance, we
assume any possible action after the measurement process
\cite{bhow13}. A general correction procedure is represented by a
trace-preserving completely positive map $\Psi$. It is used for
reconstruction of the initial system $A$ from the output system
$B$ and the measurement record. The final estimation is
then obtained by a standard measurement of $\az$ performed on the
result of correction stage. The information-theoretic disturbance
will depend on the joint probability distribution \cite{bhow13}
\begin{equation}
p(z^{\prime},z)=p(z){\,}p(z^{\prime}|z)
=\Tr\bigl(\lasf(z)\,\bro_{*}\bigr)\,p(z^{\prime}|z)
\ . \label{jpzz}
\end{equation}
This distribution characterizes correlations between the input
eigenvalue $z$ and the final estimation $z^{\prime}$. The related
conditional probability is expressed as
\begin{equation}
p(z^{\prime}|z)=\frac{1}{d_{z}}{\>}
\Tr\bigl[{\,}\lasf(z^{\prime}){\>}\Psi\circ\Phi_{\mc}\bigl(\lasf(z)\bigr)\bigr]
\>. \label{pzcz}
\end{equation}

Following \cite{bhow13}, we use the two definitions. For
$\alpha\in(0,1]$, we define R\'{e}nyi's information-theoretic
disturbance of the instrument $\mc$ as
\begin{equation}
D_{\alpha}^{(R)}(\mc,\az):=\underset{\Psi}{\min}\,R_{\alpha}(Z|Z^{\prime})
\ . \label{redis}
\end{equation}
Here, the minimization is taken over all possible TPCP maps
$\Psi$. In the case $d=2$, the measure (\ref{redis}) will be used
for $\alpha\in(0,2]$. The conditional entropy
$R_{\alpha}(Z|Z^{\prime})$ is calculated from the joint
probability distribution (\ref{jpzz}). Further, we define Tsallis'
information-theoretic disturbance
\begin{equation}
D_{\alpha}^{(T)}(\mc,\az):=\underset{\Psi}{\min}\,\hw_{\alpha}(Z|Z^{\prime})
\ , \label{tsadis}
\end{equation}
Let us discuss briefly some reasons for the above definitions. We
write (\ref{redis}) with the restriction $\alpha\in(0,1]$, since
the property (\ref{rtem2}) will be essential in the proofs.
Further, the error probability of the final estimation is written
as
\begin{equation}
q_{e}=\sum\nolimits_{z} p(e,z)
\ , \qquad
p(e,z)=\sum\nolimits_{z^{\prime}\neq{z}}{p(z^{\prime},z)}
\ . \label{qerr}
\end{equation}
As was shown in \cite{bhow13} for the non-degenerate case, the
error probability $q_{e}$ is immediately connected with the
average fidelity of correction. For non-degenerate $\az$, one has
\begin{equation}
1-q_{e}=\frac{1}{d}{\>}
\sum\nolimits_{z}{{\rff}{\bigl(\Psi\circ\Phi_{\mc}(|z\rangle\langle{z}|),|z\rangle\langle{z}|\bigr)}}
\ . \label{ferr}
\end{equation}
Recall that the Schatten $1$-norm $\|\am\|_{1}$ is
defined as the sum of all singular values $\sigma_{j}(\am)$
\cite{watrous1}. Then the fidelity between density matrices $\bro$
and $\bom$ is expressed as \cite{uhlmann76,jozsa94}
\begin{equation}
\rff(\bro,\bom)=\bigl\|\sqrt{\bro}{\,}\sqrt{\bom\vphantom{\rho}}\bigr\|_{1}^{2}
\ . \label{fidf}
\end{equation}
When the right-hand side of (\ref{ferr}) reaches $1$, the error
probability $q_{e}$ is zero and each of the quantities
(\ref{redis}) and (\ref{tsadis}) vanishes. The latter follows from
the inequalities (\ref{fn01})--(\ref{rexb1}).

\subsection{Tsallis and R\'{e}nyi formulations}\label{ssc32}

In this subsection, we will derive Tsallis and R\'{e}nyi
formulations of noise-disturbance trade-off relations. We begin
with relations that are based on the lower bounds (\ref{ovbt}) and
(\ref{ovbr}). The first result is formulated as follows.

\newtheorem{prop31}{Proposition}
\begin{prop31}\label{pan31}
Let $\mc$ be a measuring apparatus, and let
$\ax$ and $\az$ be two observables. For all $\alpha>0$ and
$\beta>0$, the Tsallis information-theoretic noise and
disturbance satisfy
\begin{equation}
N_{\alpha}^{(T)}(\mc,\ax)+D_{\beta}^{(T)}(\mc,\az)
\geq\overline{\clb}_{\alpha,\beta}^{{\,}(T)}(c)
\ , \label{bata}
\end{equation}
where the bound (\ref{ovbt}) is calculated for the characteristic
(\ref{cxzp}).
\end{prop31}

{\bf Proof.} By $\hh_{A}$, we mean the Hilbert space of the
principal quantum system. We also introduce its reference copy $C$
with the isomorphic space $\hh_{C}$. Fixing some orthonormal bases
$\{|n_{A}\rangle\}$ for $\hh_{A}$ and $\{|n_{C}\rangle\}$ for
$\hh_{C}$, one defines a maximally entangled state
\begin{equation}
|\Phi_{AC}^{+}\rangle=\frac{1}{\sqrt{d}}{\>}\sum_{n=1}^{d}{|n_{A}\rangle\otimes|n_{C}\rangle}
\ . \label{mest}
\end{equation}
For any observable $\ax_{A}\in\lsa(\hh_{A})$, we then express the
partial trace
\begin{equation}
{\Tr_{C}}{\Bigl((\pen_{A}\otimes\ax_{C})|\Phi_{AC}^{+}\rangle\langle\Phi_{AC}^{+}|\Bigr)}
=\frac{1}{d}{\>}\ax_{A}^{\dgt}
\ . \label{xdgt}
\end{equation}
Here, the operator $\ax_{A}^{\dgt}$ is transpose to $\ax_{A}$ with
respect to the prescribed basis. Hence, the so-called ``ricochet''
property holds \cite{bhow13}:
\begin{equation}
\frac{1}{d}\>|x_{A}\rangle\langle{x}_{A}|=
{\Tr_{C}}{\Bigl(\bigl(\pen_{A}\otimes|x_{C}\rangle\langle{x}_{C}|^{\dgt}\bigr)|\Phi_{AC}^{+}\rangle\langle\Phi_{AC}^{+}|\Bigr)}
\ . \label{rico}
\end{equation}
Following \cite{bhow13}, we use the fact that the two correlation
experiments defining noise and disturbance can be treated as a
single estimation producing a pair of random variables
$U=(V,V^{\prime})$. In particular, we may choose $V$ to be a copy
of $M$, while $V^{\prime}$ is the best possible estimate $\htz$
for $Z$ \cite{bhow13}. If some POVM $\bigl\{\mpi_{A}(u)\bigr\}$
with $u\in\Omega_{U}$ corresponds to the estimation of $U$, then
the conditional probabilities are expressed as
\begin{align}
p(u|x)&=\frac{1}{d_{x}}{\>}\Tr\bigl(\mpi_{A}(u){\,}\pisf_{A}(x)\bigr)
{\>}, \label{cuxpd}\\
p(u|z)&=\frac{1}{d_{z}}\>\Tr\bigl(\mpi_{A}(u){\,}\lasf_{A}(z)\bigr)
{\>}. \label{cuzpd}
\end{align}
The joint probabilities are obtained after multiplying
(\ref{cuxpd}) by $p(x)=d_{x}/d$ and (\ref{cuzpd}) by
$p(z)=d_{z}/d$, respectively. So, we write
\begin{align}
p(u,x)&=\frac{1}{d}{\>}{\Tr}{\bigl(\mpi_{A}(u){\,}\pisf_{A}(x)\bigr)}
{\>}, \label{uxpd}\\
p(u,z)&=\frac{1}{d}{\>}{\Tr}{\bigl(\mpi_{A}(u){\,}\lasf_{A}(z)\bigr)}
{\>}. \label{uzpd}
\end{align}
Due to the ``ricochet'' property (\ref{rico}) and linearity of the
transpose operation, the probabilities can be rewritten as
\begin{align}
p(u,x)&=
\Tr\Bigl(\bigl(\mpi_{A}(u)\otimes\pisf_{C}(x)^{\dgt}\bigr)|\Phi_{AC}^{+}\rangle\langle\Phi_{AC}^{+}|\Bigr)
\, , \label{rxpd}\\
p(u,z)&=
\Tr\Bigl(\bigl(\mpi_{A}(u)\otimes\lasf_{C}(z)^{\dgt}\bigr)|\Phi_{AC}^{+}\rangle\langle\Phi_{AC}^{+}|\Bigr)
\, . \label{rzpd}
\end{align}
We now consider an ensemble of mixed states $\bro_{C}(u)$ with
corresponding probabilities $p(u)$. These states and probabilities
are written as
\begin{align}
\bro_{C}(u)&=p(u)^{-1}{\,}
\Tr_{A}\Bigl(\bigl(\mpi_{A}(u)\otimes\pen_{C}\bigr)|\Phi_{AC}^{+}\rangle\langle\Phi_{AC}^{+}|\Bigr)
\, , \label{rhuc}\\
p(u)&=\Tr\Bigl(\bigl(\mpi_{A}(u)\otimes\pen_{C}\bigr)|\Phi_{AC}^{+}\rangle\langle\Phi_{AC}^{+}|\Bigr)
\, . \label{phuc}
\end{align}
We easily check that the probabilities (\ref{rxpd}) and
(\ref{rzpd}) can be represented as
\begin{align}
p(u,x)&=p(u)\,\Tr\bigl(\pisf_{C}(x)^{\dgt}\bro_{C}(u)\bigr)
\, , \label{puxd}\\
p(u,z)&=p(u)\,\Tr\bigl(\lasf_{C}(z)^{\dgt}\bro_{C}(u)\bigr)
\, . \label{puzd}
\end{align}
Hence, we have
$\Tr\bigl(\pisf_{C}(x)^{\dgt}\bro_{C}(u)\bigr)=p(x|u)$ and
$\Tr\bigl(\lasf_{C}(z)^{\dgt}\bro_{C}(u)\bigr)=p(z|u)$. Let us
apply the entropic uncertainty relation for the Tsallis entropies.
For each value of $u$, one gives
\begin{equation}
{\rh_{\alpha}}{\bigl(\ax_{C}^{\dgt};\bro_{C}(u)\bigr)}
+{\rh_{\beta}}{\bigl(\az_{C}^{\dgt};\bro_{C}(u)\bigr)}
\geq\overline{\clb}_{\alpha,\beta}^{{\,}(T)}(\tilde{c})
\ , \label{xzcp}
\end{equation}
where the parameter $\tilde{c}$ is defined as
\begin{equation}
\tilde{c}:={\max}{\Bigl\{
\|\pisf(x)^{\dgt}\,\lasf(z)^{\dgt}\|_{\infty}
:{\>}
x\in\spc(\ax),{\>}z\in\spc(\az)
\Bigr\}}
{\,}, \label{ccpp}
\end{equation}
It follows from the singular value theorem and (\ref{etcs}) that
the parameter $\tilde{c}$ coincides with (\ref{cxzp}). Multiplying
(\ref{xzcp}) by $p(u)$ and summing over all $u\in\Omega_{U}$, we
obtain
\begin{equation}
\hw_{\alpha}(X|U)+\hw_{\beta}(Z|U)\geq
\overline{\clb}_{\alpha,\beta}^{{\,}(T)}(c)
\ , \label{ucp}
\end{equation}
due to
$H_{\alpha}(X|u)={\rh_{\alpha}}{\bigl(\ax_{C}^{\dgt};\bro_{C}(u)\bigr)}$
and
$H_{\alpha}(Z|u)={\rh_{\alpha}}{\bigl(\az_{C}^{\dgt};\bro_{C}(u)\bigr)}$.
Since the property (\ref{t12t1}) holds for all $\alpha>0$, we have
\begin{align}
& N_{\alpha}^{(T)}(\mc,\ax)=\hw_{\alpha}(X|M)\geq
\hw_{\alpha}(X|M,\htz)=\hw_{\alpha}(X|U)
\ , \label{t12n}\\
& D_{\beta}^{(T)}(\mc,\az)=\hw_{\beta}(Z|\htz)\geq
\hw_{\beta}(Z|M,\htz)=\hw_{\beta}(Z|U)
\ . \label{t12d}
\end{align}
Combining (\ref{ucp}) with (\ref{t12n}) and (\ref{t12d}) completes
the proof. $\blacksquare$

In a similar manner, we will obtain a formulation in the R\'{e}nyi
case. The following statement takes place.

\newtheorem{prop32}[prop31]{Proposition}
\begin{prop32}\label{pan32}
Let $\mc$ be a measuring apparatus, and let
$\ax$ and $\az$ be two observables. When the orders $\alpha$
and $\beta$ are both in the interval $(0,1]$, the R\'{e}nyi
information-theoretic noise and disturbance satisfy
\begin{equation}
N_{\alpha}^{(R)}(\mc,\ax)+D_{\beta}^{(R)}(\mc,\az)
\geq \overline{\clb}_{\alpha,\beta}^{{\,}(R)}(c)
\ , \label{bara}
\end{equation}
where the bound (\ref{ovbr}) is calculated for the characteristic
(\ref{cxzp}). In the case ${\mathrm{dim}}(\hh_{A})=2$, the trade-off
relation (\ref{bara}) holds for $\alpha,\beta\in(0,2]$.
\end{prop32}

{\bf Proof.} Repeating the argumentation between
(\ref{mest})--(\ref{ccpp}), we merely replace (\ref{xzcp}) with
the relation
\begin{equation}
{\rrh_{\alpha}}{\bigl(\ax_{C}^{\dgt};\bro_{C}(u)\bigr)}
+{\rrh_{\beta}}{\bigl(\az_{C}^{\dgt};\bro_{C}(u)\bigr)}
\geq\overline{\clb}_{\alpha,\beta}^{{\,}(R)}(\tilde{c})
\ , \label{rxzcp}
\end{equation}
which holds for all $\alpha>0$ and $\beta>0$. Note that we have
${\rrh_{\alpha}}{\bigl(\ax_{C}^{\dgt};\bro_{C}(u)\bigr)}=R_{\alpha}(X|u)$
and
${\rrh_{\beta}}{\bigl(\az_{C}^{\dgt};\bro_{C}(u)\bigr)}=R_{\beta}(Z|u)$.
Multiplying (\ref{rxzcp}) by $p(u)$ and summing over all
$u\in\Omega_{U}$, we obtain
\begin{equation}
R_{\alpha}(X|U)+R_{\beta}(Z|U)\geq
\overline{\clb}_{\alpha,\beta}^{{\,}(R)}(c)
\ . \label{rucp}
\end{equation}
Similarly to (\ref{t12n}) and (\ref{t12d}), we write the following
relations. When both the orders $\alpha$ and $\beta$ lie in the
range $(0,1]$, the property (\ref{rtem2}) leads to
\begin{align}
& N_{\alpha}^{(R)}(\mc,\ax)=R_{\alpha}(X|M)\geq
R_{\alpha}(X|M,\htz)=R_{\alpha}(X|U)
\ , \label{r12n}\\
& D_{\beta}^{(R)}(\mc,\az)=R_{\beta}(Z|\htz)\geq
R_{\beta}(Z|M,\htz)=R_{\beta}(Z|U)
\ . \label{r12d}
\end{align}
If $d=2$, these relations holds for $\alpha,\beta\in(0,2]$.
Combining (\ref{rucp}) with (\ref{r12n}) and (\ref{r12d})
completes the proof. $\blacksquare$

Propositions \ref{pan31} and \ref{pan32} are respectively the Tsallis
and R\'{e}nyi formulations of relations for noise and disturbance.
In a certain sense, they are an extension of the
noise-disturbance relation given in \cite{bhow13}. In our
notation, the information-theoretic relation of the paper \cite{bhow13}
is written as
\begin{equation}
N_{1}(\mc,\ax)+D_{1}(\mc,\az)
\geq -2\ln{c}
\ . \label{stnd}
\end{equation}
The authors of \cite{bhow13} defined the information-theoretic
noise and disturbance in terms of the standard conditional
entropy. So, we left out superscripts in the formula (\ref{stnd}).
Each of the definitions (\ref{renos}) and (\ref{tsanos}) leads to
the standard information-theoretic noise in the limit
$\alpha\to1$. In the same limit, both the definitions
(\ref{redis}) and (\ref{tsadis}) gives the standard
information-theoretic disturbance of \cite{bhow13}. The bounds
(\ref{ovbt}) and (\ref{ovbr}) are not always $c$-optimal in
general. Moreover, for $\alpha=\beta=1$ these bounds do not
coincide with the Maassen--Uffink bound. Thus, the relations
(\ref{bata}) and (\ref{bara}) do not lead to (\ref{stnd}) in the
case $\alpha=\beta=1$. We shall now derive such a direct
extension. It is based on the entropic bound (\ref{gntl}).

\newtheorem{prop33}[prop31]{Proposition}
\begin{prop33}\label{pan33}
Let $\mc$ be a measuring apparatus, and let
$\ax$ and $\az$ be two observables.
If $\alpha>0$ and $\beta>0$ obey $1/\alpha+1/\beta=2$, then
\begin{equation}
N_{\alpha}^{(T)}(\mc,\ax)+D_{\beta}^{(T)}(\mc,\az)
\geq{\ln_{\mu}}{\bigl(c^{-2}\bigr)}
\ , \label{tpil0}
\end{equation}
where $\mu=\max\{\alpha,\beta\}$ and
the characteristic $c$ is defined by (\ref{cxzp}).
\end{prop33}

{\bf Proof.} The argumentation can be followed like the proof of
Proposition \ref{pan31}. For each $u$, combining (\ref{puxd}) and (\ref{puzd})
with (\ref{gntl}) finally gives
\begin{equation}
{\rh_{\alpha}}{\bigl(\ax_{C}^{\dgt};\bro_{C}(u)\bigr)}
+{\rh_{\beta}}{\bigl(\az_{C}^{\dgt};\bro_{C}(u)\bigr)}
\geq{\ln_{\mu}}{\bigl(c^{-2}\bigr)}
\ , \label{xzzcp}
\end{equation}
where $\mu=\max\{\alpha,\beta\}$ and $1/\alpha+1/\beta=2$. Then we
complete the argumentation similarly to the proof of
Proposition \ref{pan31}. $\blacksquare$

As a particular case of (\ref{tpil0}), we have the noise-disturbance relation (\ref{stnd})
derived in \cite{bhow13}. Thus, our result (\ref{tpil0}) is an immediate
extension of (\ref{stnd}). A final comment concerns possible
R\'{e}nyi's formulation based on (\ref{gnrl}). Here, the concavity
and related properties are crucial. If the dimensionality is not
prescribed, the property (\ref{rtem2}) can be accepted only for
$\alpha\in(0,1]$. Combining the latter with $1/\alpha+1/\beta=2$
gives $\alpha=\beta=1$. With (\ref{gnrl}), therefore, we could
reach no more than (\ref{stnd}). In the two-dimensional case, we can get a
little extension. Here, non-trivial observables are certainly
non-degenerate. For $d=2$, we have
\begin{equation}
N_{\alpha}^{(R)}(\mc,\ax)+D_{\beta}^{(R)}(\mc,\az)
\geq -2\ln{c}
\ , \label{stndr1}
\end{equation}
where $1/\alpha+1/\beta=2$ and $\alpha,\beta\in(0,2]$. A search
for tightest bounds remains open in general. Novel uncertainty
relations would lead to new trade-off relations for noise and
disturbance.

\section{Conclusions}\label{sec4}

We have obtained trade-off relations for noise and
disturbance in terms of the R\'{e}nyi and Tsallis
information-theoretic measures. Our work is a further development
of the approach originally proposed in \cite{bhow13}. As was shown
in several cases, the use of generalized entropies may give new
possibilities in analyzing statistical data. The presented
information-theoretic measures of noise and disturbance are based
on the conditional R\'{e}nyi and Tsallis entropies. Introduced
measures were motivated with the use of important properties of
the conditional entropies. In particular, relations between the
conditional entropies and the error probability were essential. We
utilized several formulations of entropic uncertainty relations
for a pair of observables. These formulations lead to trade-off
relations for introduced measures of noise and disturbance. The
scope of obtained results also depends on concavity properties of
the considered entropies. In this regard, the R\'{e}nyi
formulation turns out to be somewhat restricted. In the
noise-disturbance relations (\ref{bata}) and (\ref{bara}), the
entropic parameters do not satisfy any constraint. We only specify
an interval, in which the parameters should range. When the
entropic parameters obey a certain constraint, we can use entropic
bounds of the Maassen--Uffink type. Hence, we have obtained the
noise-disturbance relations (\ref{tpil0}) and (\ref{stndr1}).

\acknowledgments

The author is grateful to
Francesco Buscemi for useful correspondence and to anonymous
referees for valuable comments.


\begin{thebibliography}{100}

\bibitem{heisenberg}
W.~Heisenberg (1927), {\it \"{U}ber den anschaulichen Inhalt der
quanten theoretischen Kinematik und Mechanik}, Zeitschrift f\"{u}r
Physik {\bf 43}, 172--198.

\bibitem{lahti}
P.~Busch, T.~Heinonen, and P.~J.~Lahti (2007), {\it Heisenberg's uncertainty principle}, Phys. Rep. {\bf 452}, 155--176.

\bibitem{ww10}
S.~Wehner and A.~Winter (2010), {\it Entropic uncertainty relations -- a survey}, New J. Phys. {\bf 12}, 025009.

\bibitem{brud11}
I.~Bia{\l}ynicki-Birula and {\L}.~Rudnicki (2011), {\it Entropic uncertainty
relations in quantum physics}, In K.~D.~Sen, ed., Statistical
Complexity, 1--34, Springer (Berlin).

\bibitem{hall}
M.~J.~W.~Hall (1999), {\it Universal geometric approach to uncertainty, entropy, and information}, Phys. Rev. A {\bf 59}, 2602--2615.

\bibitem{kennard}
E.~H.~Kennard (1927), {\it Zur Quantenmechanik einfacher Bewegungstypen}, Zeitschrift f\"{u}r
Physik {\bf 44}, 326--352.

\bibitem{robert}
H.~P.~Robertson (1929), {\it The uncertainty principle}, Phys. Rev. {\bf 34}, 163--164.

\bibitem{deutsch}
D.~Deutsch (1983), {\it Uncertainty in quantum measurements}, Phys. Rev. Lett. {\bf 50}, 631--633.

\bibitem{maass}
H.~Maassen and J.~B.~M.~Uffink (1988), {\it Generalized entropic uncertainty relations}, Phys. Rev. Lett. {\bf 60}, 1103--1106.

\bibitem{BCCRR10}
M.~Berta, M.~Christandl, R.~Colbeck, J.~M.~Renes, and R.~Renner (2010), {\it The uncertainty principle in the presence of quantum memory},
Nature Phys. {\bf 6}, 659--662.

\bibitem{ccyz12}
P.~J.~Coles, R.~Colbeck, L.~Yu, and M.~Zwolak (2012), {\it Uncertainty relations from simple entropic properties}, Phys. Rev. Lett. {\bf 108}, 210405.

\bibitem{mdsft13}
M.~M\"{u}ller-Lennert, F.~Dupuis, O.~Szehr, S.~Fehr, and
M.~Tomamichel (2013), {\it On quantum R\'{e}nyi entropies: A new
generalization and some properties}, J. Math. Phys. {\bf 54},
122203.

\bibitem{tomamichel15}
M.~Tomamichel (2015), {\it Quantum information processing with finite resources}, e-print arXiv:1504.00233 [quant-ph].

\bibitem{ozawa03}
M.~Ozawa (2003), {\it Uncertainty principle for quantum instruments and computing}, Int. J. Quantum Inf. {\bf 1}, 569--588.

\bibitem{ozawa033}
M.~Ozawa (2003), {\it Universally valid reformulation of the Heisenberg
uncertainty principle on noise and disturbance in measurement},
Phys. Rev. A {\bf 67}, 042105.

\bibitem{hall04}
M.~J.~W.~Hall (2004), {\it Prior information: How to circumvent
the standard joint-measurement uncertainty relation}, Phys. Rev. A
{\bf 69}, 052113.

\bibitem{werner04}
R.~F.~Werner (2004), {\it The uncertainty relation for joint
measurement of position and momentum}, Quantum Inf. Comput. {\bf
4}, 546--562.

\bibitem{wsu11}
Y.~Watanabe, T.~Sagawa, and M.~Ueda (2011), {\it Uncertainty
relation revisited from quantum estimation theory}, Phys. Rev. A
{\bf 84}, 042121.

\bibitem{blw13}
P.~Busch, P.~Lahti, and R.~F.~Werner (2013), {\it Proof of
Heisenberg's error-disturbance relation}, Phys. Rev. Lett. {\bf
111}, 160405.

\bibitem{yuoh13}
X.-M.~Lu, S.~Yu, K.~Fujikawa, and C.~H.~Oh (2014), {\it Improved
error-tradeoff and error-disturbance relations in terms of
measurement error components}, Phys. Rev. A {\bf 90}, 042113.

\bibitem{mans14}
P.~Mandayam and M.~D.~Srinivas (2014), {\it A disturbance tradeoff
principle for incompatible quantum observables}, arXiv:1402.7311
[quant-ph].

\bibitem{kboe14}
F.~Kaneda, S.-Y.~Baek, M.~Ozawa, and K.~Edamatsu (2014), {\it
Experimental  test of error-disturbance uncertainty relations by
weak measurement}, Phys. Rev. Lett. {\bf 112}, 020402.

\bibitem{blw14}
P.~Busch, P.~Lahti, and R.~F.~Werner (2014), {\it Comments on
``Experimental test of error-disturbance uncertainty relations by
weak measurement''}, arXiv:1403.0367 [quant-ph].

\bibitem{bhow13}
F.~Buscemi, M.~J.~W.~Hall, M.~Ozawa, and M.~M.~Wilde (2014), {\it Noise and
disturbance in quantum measurements: An information-theoretic
approach}, Phys. Rev. Lett. {\bf 112}, 050401.

\bibitem{cofu13}
P.~J.~Coles and F.~Furrer (2015), {\it State-dependent approach to
entropic measurement-disturbance relations}, Phys. Lett. A {\bf
379}, 105--112.

\bibitem{tomamichel11}
M.~Tomamichel and R.~Renner (2011), {\it Uncertainty relation for smooth entropies}, Phys. Rev. Lett. {\bf 106}, 110506.

\bibitem{ngbw12}
H.~Y.~N.~Ng, M.~Berta, and S.~Wehner (2012), {\it Min-entropy
uncertainty relation for finite-size cryptography}, Phys. Rev. A
{\bf 86}, 042315.

\bibitem{peresq}
A.~Peres (1993), {\it Quantum Theory: Concepts and Methods}, Kluwer (Dordrecht).

\bibitem{prug77}
E.~Prugove\u{c}ki (1977), {\it Information-theoretical aspects of quantum measurement}, Int. J. Theor. Phys. {\bf 16}, 321--331.

\bibitem{busch91}
P.~Busch (1991), {\it Informationally complete sets of physical quantities}, Int. J. Theor. Phys. {\bf 30}, 1217--1227.

\bibitem{rbsc04}
J.~M.~Renes, R.~Blume-Kohout, A.~J.~Scott, and C.~M.~Caves (2004), {\it Symmetric
informationally complete quantum measurements}, J. Math. Phys. {\bf
45}, 2171--2180.

\bibitem{rastmub}
A.~E.~Rastegin (2013), {\it Uncertainty relations for MUBs and SIC-POVMs in terms of generalized entropies}, Eur. Phys. J. D {\bf 67}, 269.

\bibitem{dabo14}
M.~Dall'Arno, F.~Buscemi, and M.~Ozawa (2014), {\it Tight bounds
on accessible information and informational power}, J. Phys. A:
Math. Theor. {\bf 47}, 235302.

\bibitem{dord13}
J.~Dressel and A.~N.~Jordan (2013), {\it Quantum instruments as a foundation for both states and observables}, Phys. Rev. A {\bf 88}, 022107.

\bibitem{nielsen}
M.~A.~Nielsen and I.~L.~Chuang (2000), {\it Quantum Computation and
Quantum Information}, Cambridge University Press (Cambridge).

\bibitem{wilde13}
M.~M.~Wilde (2013), {\it Quantum Information Theory}, Cambridge University Press (Cambridge).

\bibitem{bengtsson}
I.~Bengtsson and K.~\.{Z}yczkowski (2006), {\it Geometry of Quantum States: An
Introduction to Quantum Entanglement}, Cambridge University Press (Cambridge).

\bibitem{renyi61}
A.~R\'{e}nyi (1961), {\it On measures of entropy and information}. In J.~Neyman,
ed., Proceedings of 4th Berkeley symposium on mathematical
statistics and probability, vol. I, 547--561, University of
California Press (Berkeley).

\bibitem{zycz}
K.~\.{Z}yczkowski (2003), {\it R\'{e}nyi extrapolation of Shannon entropy}, Open Sys. Inf. Dyn. {\bf 10}, 297--310;
corrigendum in arXiv:quant-ph/0305062v2.

\bibitem{ja04}
P.~Jizba and T.~Arimitsu (2004), {\it The world according to R\'enyi: thermodynamics of multifractal systems}, Ann. Phys. {\bf 312}, 17--59.

\bibitem{ben78}
M.~Ben-Bassat and J.~Raviv (1978), {\it R\'{e}nyi's entropy and error probability}, IEEE Trans. Inf. Theory {\bf 24}, 324--331.

\bibitem{tsallis}
C.~Tsallis (1988), {\it Possible generalization of Boltzmann--Gibbs statistics}, J. Stat. Phys. {\bf 52}, 479--487.

\bibitem{rprz12}
W.~Roga, Z.~Pucha{\l}a, {\L}.~Rudnicki, and K.~\.{Z}yczkowski (2013),
{\it Entropic trade-off relations for quantum operations}, Phys. Rev. A
{\bf 87}, 032308.

\bibitem{rast13a}
A.~E.~Rastegin (2013), {\it Unified-entropy trade-off relations for a single quantum channel}, J. Phys. A: Math. Theor. {\bf 46}, 285301.

\bibitem{CT91}
T.~M.~Cover and J.~A.~Thomas (1991), {\it Elements of Information Theory}, John Wiley {\&} Sons (New York).

\bibitem{sf06}
S.~Furuichi (2006), {\it Information theoretical properties of Tsallis entropies}, J. Math. Phys. {\bf 47}, 023302.

\bibitem{rastkyb}
A.~E.~Rastegin (2012), {\it Convexity inequalities for estimating generalized conditional entropies from below}, Kybernetika {\bf 48}, 242--253.

\bibitem{ZD70}
Z.~Dar\'{o}czy (1970), {\it Generalized information functions}, Inform. Control {\bf 16}, 36--51.

\bibitem{rastit}
A.~E.~Rastegin (2015), {\it Further results on generalized conditional entropies}, RAIRO--Theor. Inf. Appl. {\bf 49}, 67--92.

\bibitem{tma12}
A.~Teixeira, A.~Matos, and L.~Antunes (2012), {\it Conditional R\'{e}nyi entropies}, IEEE Trans. Inf. Theory {\bf 58}, 4273--4277.

\bibitem{cch97}
C.~Cachin (1997), {\it Entropy measures and unconditional security in cryptography}, PhD Thesis, Swiss Federal Institute of Technology (Z\"{u}rich).

\bibitem{Kam98}
R.~Kamimura (1998), {\it Minimizing $\alpha$-information for generalization and interpretation}, Algorithmica, {\bf 22}, 173--197.

\bibitem{EP04}
D.~Erdogmus and J.~C.~Principe (2004), {\it Lower and upper bounds for misclassification probability based on R\'{e}nyi's information}, J.
VLSI Signal Process. {\bf 37}, 305--317.

\bibitem{renyi67a}
A.~R\'{e}nyi (1967), {\it Statistics and information theory}, Stud. Sci. Math. Hung. {\bf 2}, 249--256.

\bibitem{rw05}
R.~Renner and S.~Wolf (2005), {\it Simple and tight bounds for information
reconciliation and privacy amplification}, In B.~Roy, ed.,
Advances in Cryptology --- ASIACRYPT 2005, Lecture Notes in Computer Science, vol. 3788, 199--216, Springer (Berlin).

\bibitem{gpy09}
L.~Golshani, E.~Pasha, and G.~Yari (2009), {\it Some properties of R\'{e}nyi
entropy and R\'{e}nyi entropy rate}, Information Sciences {\bf
179}, 2426--2433.

\bibitem{FM94}
M.~Feder and N.~Merhav (1994), {\it Relations between entropy and error probability}, IEEE Trans. Inform. Theory {\bf 40}, 259--266.

\bibitem{AP67}
A.~Perez (1967), {\it Information-theoretic risk estimates in statistical decision}, Kybernetika {\bf 3}, 1--21.

\bibitem{fano61}
R.~M.~Fano (1961), {\it Transmission of Information: A Statistical Theory of Communications}, MIT Press and John Wiley \&\ Sons (New York).

\bibitem{hv10}
S.-W.~Ho and S.~Verd\'{u} (2011), {\it On the interplay between
conditional entropy and error probability}, IEEE Trans. Inf.
Theory {\bf 56}, 5930--5942.

\bibitem{zozor13}
S.~Zozor, G.~M.~Bosyk, and M.~Portesi (2013), {\it On a generalized entropic
uncertainty relations in the case of the qubit}, J. Phys. A: Math.
Theor. {\bf 46}, 465301.

\bibitem{zozor1311}
S.~Zozor, G.~M.~Bosyk, and M.~Portesi (2014), {\it General
entropic-like uncertainty relations for $N$-level systems}, J.
Phys. A: Math. Theor. {\bf 47}, 495302.

\bibitem{watrous1}
J.~Watrous, {\it Theory of Quantum Information}, a draft of book,
University of Waterloo (Waterloo).
http://cs.uwaterloo.ca/{\textasciitilde}watrous/TQI/

\bibitem{sal1}
M.~Salicr\'{u}, M.~L.~Men\'{e}ndez, D.~Morales, and L.~Pardo (1993), {\it Asymptotic distribution of $(h,\phi)$-entropies},
Communications in Statistics -- Theory and Methods {\bf 22}, 2015--2031.

\bibitem{sal2}
M.~L.~Men\'{e}ndez, D.~Morales, L.~Pardo, and M.~Salicr\'{u}
(1997), {\it $(h,\phi)$-entropy differential metric}, Applications
of Mathematics {\bf 42}, 81--98.

\bibitem{birula06}
I.~Bia{\l}ynicki-Birula (2006), {\it Formulation of the uncertainty relations in terms of the R\'{e}nyi entropies}, Phys. Rev. A {\bf 74}, 052101.

\bibitem{rast104}
A.~E.~Rastegin (2011), {\it Entropic uncertainty relations for extremal unravelings of super-operators}, J. Phys. A: Math. Theor.
{\bf 44}, 095303.

\bibitem{rast12num}
A.~E.~Rastegin (2012), {\it Number-phase uncertainty relations in terms of generalized entropies}, Quantum Inf. Comput. {\bf 12}, 0743--0762.

\bibitem{rast12quasi}
A.~E.~Rastegin (2012), {\it Entropic uncertainty relations and quasi-Hermitian operators}, J. Phys. A: Math. Theor. {\bf 45}, 444026.

\bibitem{bosyk13a}
G.~M.~Bosyk, M.~Portesi, F.~Holik, A.~Plastino (2013), {\it On the
connection between complementarity and uncertainty principles in
the Mach--Zehnder interferometric setting}, Phys. Scr. {\bf 87},
065002.

\bibitem{prz13}
Z.~Pucha{\l}a, {\L}.~Rudnicki, and K.~\.{Z}yczkowski (2013), {\it Majorization entropic uncertainty relations}, J. Phys. A: Math. Theor. {\bf 46}, 272002.

\bibitem{fgg13}
S.~Friedland, V.~Gheorghiu, and G.~Gour (2013), {\it Universal uncertainty relations}, Phys. Rev. Lett. {\bf 111}, 230401.

\bibitem{rpz14}
{\L}.~Rudnicki, Z.~Pucha{\l}a, and K.~\.{Z}yczkowski (2014), {\it Strong majorization entropic uncertainty relations}, Phys. Rev. A {\bf 89}, 052115.

\bibitem{uhlmann76}
A.~Uhlmann (1976), {\it The transition probability in the state space of a *-algebra}. Rep. Math. Phys. {\bf 9}, 273--279.

\bibitem{jozsa94}
R.~Jozsa (1994), {\it Fidelity for mixed quantum states}, J. Mod. Opt. {\bf 41}, 2315--2323.


\end{thebibliography}
\end{document}